\documentclass[twocolumn,preprintnumbers]{revtex4}
\usepackage{epsf}
\usepackage[dvips]{color}
\newcommand{\ket}[1]{| #1 \rangle}

\begin{document}
\preprint{KUNS-2364}
\title{$^6$He-triton cluster states in $^9$Li}
\author{Yoshiko Kanada-En'yo}
\affiliation{Department of Physics, Kyoto University, Kyoto 606-8502, Japan}

\author{Tadahiro Suhara}
\affiliation{Yukawa Institute for Theoretical Physics, Kyoto University,
Kyoto 606-8502, Japan}

\begin{abstract}
Cluster states in $^9$Li are investigated with calculations of a $^6$He-$t$ cluster model.
Results suggest $^6$He-$t$ cluster states near the $^6$He-$t$ threshold energy.
These states construct  a $K^\pi=1/2^-$ band and their neutron configuration is similar to that of the $K^\pi=0^+_2$ band in $^{10}$Be. 
\end{abstract}
\maketitle

\noindent

\section{Introduction} \label{ref:introduction}
It has been revealed that cluster structures appear in light unstable nuclei as well as
light stable nuclei.
In the recent progress of experimental and theoretical 
researches on unstable nuclei, various exotic cluster states have
been discovered in neutron-rich nuclei where valence neutrons play an important role. 

For instance, a variety of cluster structures have been found in neutron-rich Be isotopes.
Many of low-lying states of Be isotopes are understood in a molecular orbital picture where 
a Be nucleus is considered to consist of 2 $\alpha$'s and 
valence neutron(s) in molecular orbitals around the $2\alpha$ core \cite{OKABE,SEYA,OERTZEN,OERTZENa,ARAI,Dote:1997zz,ENYObe10,ITAGAKI,OGAWA,Arai01,AMDrev,KanadaEn'yo:2003ue,Oertzen-rev}. 
In highly excited states near the He+He threshold energy of $^{10}$Be and $^{12}$Be, 
well-developed cluster states have been suggested 
experimentally and theoretically \cite{ENYObe10,KanadaEn'yo:2003ue,Oertzen-rev,FREER,SAITO04,Curtis:2004wr,Bohlen:2007qx,Ito:2003px,Arai:2004yf,Ito:2008zza}.
Those developed cluster states show two-body 
cluster structures such as $^6$He+$^4$He in $^{10}$Be and $^6$He+$^6$He in $^{12}$Be.
Also in the neighboring nuclei, $^{14}$Be and $^{15}$B, developed two-body cluster states 
at high excitation energy have been suggested theoretically \cite{KanadaEn'yo:2002ay}.

Excitation energies of those developed two-body cluster states 
in neutron-rich nuclei can be understood systematically 
from the point of view of Ikeda's threshold rule which suggests appearances of developed cluster states 
near the corresponding threshold energy \cite{Ikeda68}. 
From the Ikeda's threshold rule, we expect possible appearance of $^6$He+$t$ cluster states near the $^6$He+$t$  threshold energy in excited states of $^9$Li.
It is a challenging issue in physics of unstable nuclei to search for such resonances of 
two neutron-rich clusters.

In our previous study with the quadrupole deformation ($\beta$-$\gamma$) constraint 
in a framework of antisymmetrized molecular dynamics (AMD), 
we have shown an indication of a largely deformed state having a $^6$He+$t$ structure
in excited states of $^9$Li \cite{Suhara:2009jb}.
The energy surface on the $\beta$-$\gamma$ plane shows a shallow local minimum in the large prolate 
region. The intrinsic wave function at the local minimum
has a developed $^6$He+$t$ cluster feature and it shows 
an elongate prolate shape of the neutron density  
which is analogous to the neutron structure of the $^{10}$Be$(0^+_2)$ 
having a developed $^6$He+$^4$He cluster feature.
Note that the $^6$He+$t$ and $^6$He+$^4$He cluster structures in these states are not weak-coupling cluster states
but rather strong coupling ones 
where a $t$ or a $^4$He cluster is sitting 
on the head of a deformed $^6$He cluster. 
According to the previous calculation, the excitation energy of the $^6$He+$t$ cluster state in $^9$Li 
is expected to be 2 MeV higher than that of the $^{10}$Be$(0^+_2)$, and hence, it might be 
a state above the $^6$He+$t$ threshold.


To investigate $^6$He+$t$-cluster states near the $^6$He+$t$ threshold in $^9$Li, 
we perform GCM calculations with a $^6$He+$t$ cluster model.
We first show the energy levels and $E2$ strengths 
obtained by the GCM calculations in a bound state approximation.
Then we discuss stability and spectra of the resonances by an analysis using 
a pseudo potential method. 

This paper is organized as follows.
In section \ref{sec:formulation}, we explain the formulation of 
the present calculations. We show the calculated results in section \ref{sec:results}
and finally give a summary in section \ref{sec:summary}.

\section{Formulation}\label{sec:formulation}
To incorporate $^6$He+$t$ resonance features, we use
Bloch-Brink (BB) \cite{brink66} $^6$He+$t$ cluster wave functions  having various inter-cluster distances 
and superpose them. The $^6$He and $t$ cluster wave functions are written 
by harmonic oscillator (HO) shell-model wave functions 
localized at $(0,0,-d/3)$ and $(0,0,+2d/3)$, respectively. Here $d$ indicates the
distance parameter, which is treated as a generator coordinate in the superposition. 
The width parameter $\nu=1/2b^2$ is common for $^6$He and $t$ clusters.

A $t$ cluster is given by the $(0s)_\pi(0s)_\nu^2$ configuration shell-model wave function.
A configuration for a $^6$He cluster is assumed to be 
$(0s)_\pi^2 (0s)_\nu^2 (0p)_\nu^2$ which indicates an $\alpha$ cluster and 
two valence neutrons in $p$ shell. For $p$-shell neutron configurations, we choose
$(p_{3/2})^2$ coupling to the total angular momentum $J_{12}=0,2$ in the $j$-$j$ coupling scheme, 
and also the total intrinsic spin $S_{12}=0$ configurations 
in the $l$-$s$ coupling scheme.
Namely, the neutron configurations are given by 
$\ket{p_z, n\uparrow}\ket{p_z, n\downarrow}$ and 
$\ket{p_{(+)}, n\uparrow}\ket{p_{(-)}, n\downarrow}$, and their rotated states.
Here $p_{(+)}, p_z, p_{(-)}$ stand for $l_z=+1,0,-1$ ($z$-component of orbital angular momentum) states in $p$ shell, respectively. 

\begin{figure}[th]
\epsfxsize=6 cm
\centerline{\epsffile{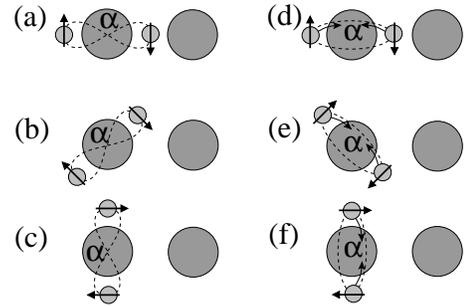}}
\caption{Schematic figures for configurations $(a)$-$(f)$ of a $^6$He cluster in the BB
$^6$He+$t$ and $^6$He+$\alpha$ cluster models.}
\label{fig:he6-t}
\end{figure}

Then, the $^6$He+$t$ cluster wave functions projected onto parity and 
total-angular-momentum eigen states are written as,
\begin{eqnarray}
&&P^{J\pm}_{MK} \ket{\Phi_{\tau}(d)}=P^{J\pm}_{MK} {\cal A}\left\{ \ket{\psi_{1\tau}(-\frac{d}{3})}\ket{\psi_{2\tau}(-\frac{d}{3})} \right. \nonumber\\
&&\times \ket{\phi(-\frac{d}{3})p\uparrow}
\ket{\phi(-\frac{d}{3}) p\downarrow}
\ket{\phi(-\frac{d}{3}) n\uparrow}
\ket{\phi(-\frac{d}{3}) n\downarrow}  \nonumber\\ 
&& \times \left. \ket{\phi(\frac{2d}{3}) p\uparrow}
\ket{\phi(\frac{2d}{3})n\uparrow}
\ket{\phi(\frac{2d}{3})n\downarrow} \right\}.
\end{eqnarray}
Here $\phi(-d/3)$ and $\phi(2d/3)$  are shifted $0s$ wave functions localized at $(0,0,-d/3)$
and $(0,0,2d/3)$, respectively. 
$\ket{\psi_{1\tau}(-\frac{d}{3})}$ and $\ket{\psi_{2\tau}(-\frac{d}{3})}$
indicate single-particle states for two valence neutrons, and given by the $p$-shell orbits 
shifted at $(0,0,-d/3)$. 
Six configurations labeled by $\tau=\{a,b,c,d,e,f \}$, which are illustrated in Fig.~\ref{fig:he6-t}, are used to 
describe valence neutron configurations of a $^6$He cluster. 
In the configuration $\tau=a$, two neutron orbitals 
$\ket{\psi_{1a}(-d/3)}$ and $\ket{\psi_{2a}(-d/3)}$ are chosen to be 
$\ket{p_z, n\uparrow}_{-d/3}$ and $\ket{p_z, n\downarrow}_{-d/3}$ that are $p$ orbits around $(0,0,-d/3)$. 
We define a rotational operator $R(\theta,-d/3)$ for the rotation around the point $(0,0,-d/3)$ with respect to
the vector $(1,0,0)$. Then, two neutron orbitals for the configurations $(b)$ and $(c)$ can be written
as  
\begin{eqnarray}
&\ket{\psi_{1\tau}(-d/3)}=R(\theta,-d/3)\ket{p_z, n\uparrow}_{-d/3},\\
&\ket{\psi_{2\tau}(-d/3)}=R(\theta,-d/3)\ket{p_z, n\downarrow}_{-d/3},
\end{eqnarray}
with $\theta=\pi/4$ and $\theta=\pi/2$, respectively. 
Similarly, two neutrons orbitals for the configuration $(f)$ are 
\begin{eqnarray}
&\ket{\psi_{1f}(-d/3)}=\ket{p_{(+)}, n\uparrow}_{-d/3},\\
&\ket{\psi_{2f}(-d/3)}=\ket{p_{(-)}, n\downarrow}_{-d/3},
\end{eqnarray}
and
those for the configurations $(d)$ and $(e)$ can be written as 
\begin{eqnarray}
&\ket{\psi_{1\tau}(-d/3)}=R(\theta,-d/3)\ket{p_{(+)}, n\uparrow}_{-d/3},\\
&\ket{\psi_{2\tau}(-d/3)}=R(\theta,-d/3)\ket{p_{(-)}, n\downarrow}_{-d/3},
\end{eqnarray}
with $\theta=\pi/2$ and  $\theta=\pi/4$,
respectively.
Configurations $(a)$, $(b)$, and $(c)$ correspond
to $|L_{12}=0,2\rangle \otimes |S_{12}=0\rangle$ states of a $^6$He cluster in the $l$-$s$ coupling scheme,
while $(d)$, $(e)$, and $(f)$ indicate the $(p_{3/2})^2$ configurations of a $^6$He cluster in the $j$-$j$ coupling scheme.
Here $L_{12}$ stands for the magnitude of total orbital angular momentum of two valence neutrons. 
The projected $^6$He+$t$ cluster wave functions for these six configurations cover 
all $0^+$ and $2^+$ states in $p$-shell configurations of a $^6$He cluster and spin-up and spin-down configurations for a 
triton cluster.

We superpose $^6$He-$t$ cluster wave functions,
\begin{equation} \label{eq:gcm}
\ket{\Psi^{J_k\pm}_M}=\sum_d \sum_{\tau,K} c^{(k)}_{d,\tau, K} P^{J\pm}_{MK} \ket{\Phi_{\tau}(d)}, 
\end{equation}
where coefficients are determined by diagonalizing norm and Hamiltonian matrices. 
This corresponds to a calculation of a generator coordinate method GCM with the generator coordinate $d$.  

We can also perform a GCM calculation for $^{10}$Be with a $^6$He+$\alpha$ cluster model
in a similar way by replacing a $t$ cluster with
an $\alpha$ cluster.
 
As already mentioned, the cluster structures suggested in $^{9}$Li and $^{10}$Be near the
$^6$He+$t$ and $^6$He+$^4$He threshold energies may be not weak-coupling cluster states
but contain strong coupling cluster components where the inter-cluster motion couples strongly to the orientation of 
a deformed $^6$He cluster, i.e., valence neutron configurations. In preceding works \cite{OERTZEN,OERTZENa,ARAI,Dote:1997zz,ENYObe10,ITAGAKI,OGAWA,Arai01,Ito:2003px}, 
the $K^\pi=0^+_2$ band of $^{10}$Be is considered to be a molecular orbital state having two valence neutrons 
in the molecular $\sigma$ orbital around the $2\alpha$ cluster core. 
In the case that $\alpha$-$\alpha$ distance is moderate, the configuration $(a)$ in Fig.~\ref{fig:he6-t} 
corresponds to the molecular orbital $\sigma^2$ state as the valence neutron orbital has a nodal structure
along the $z$-axis due to the antisymmetrization with neutrons in $\alpha$ clusters, and it 
is nothing but the strong coupling cluster structure. 
On the other hand, in the asymptotic region that the inter-cluster distance is far enough, a two-cluster system 
should become a weak coupling state where a $^6$He cluster has a certain spin and parity $J^\pi$ which weakly couple with
inter-cluster motion. In the present framework, the transition between the strong coupling 
regime to the weak coupling regime 
is taken into account by the 
linear combination of configurations $(a)$-$(f)$ projected onto the total angular momentum eigen states.

Practically, we express a configuration of a BB wave function by using a single AMD wave function
which is given by a Slater determinant of single-particle Gaussian wave packets.
A general form of AMD wave functions is described, for example, in Refs.~\cite{AMDrev,ENYObc}.

\section{Results} \label{sec:results}
\subsection{Effective nuclear forces and parameters} \label{subsec:effectiveforce}
The effective Hamiltonian is  
\begin{equation}
H_{\rm eff}=\sum_i t_i +\sum_{i<j} v_{ij}, 
\end{equation}
where the first term is kinetic term and the second term for two-body interactions consists of 
effective nuclear forces and Coulomb force. 
The adopted effective nuclear forces are 
the Volkov No.2 force ~\cite{Volkov} for finite-range central force
and the spin-orbit term of the G3RS force~\cite{LS} for spin-orbit force.
These nuclear forces have been used in many works on structures 
of $^{10}$Be and $^9$Li \cite{Dote:1997zz,ITAGAKI,Suhara:2009jb}.
The interaction parameters are $(b=h=0.125, m=0.60)$ for the Volkov No.2 force and
$u_{I}=-u_{II}=1600$ MeV for the strength of the spin-orbit force. These are the same as those
used in Refs.~\cite{Suhara:2009jb,Suhara:2010ww}.
Coulomb force is approximated by seven-range Gaussians. 
 
As for the width parameter $\nu$ of the HO shell-model wave functions for $^6$He, $t$, and $^4$He clusters,
the parameter $\nu=0.235$ fm$^{-2}$ used in Refs.~\cite{Suhara:2009jb,Suhara:2010ww} is adopted.
The generator coordinate $d$ in the GCM calculations is taken to be $d=1,2,\cdots, 8$ fm. 
The truncation of the coordinate $d\le 8$ fm in the GCM calculation corresponds to a bound-state approximation. 
To see resonance features and coupling with continuum states, we 
also take a larger model space, $d=1,2,\cdots, 15$ fm.

\subsection{Energy levels of $^9$Li and $^{10}$Be}

\begin{figure}[th]
\epsfxsize=6 cm
\centerline{\epsffile{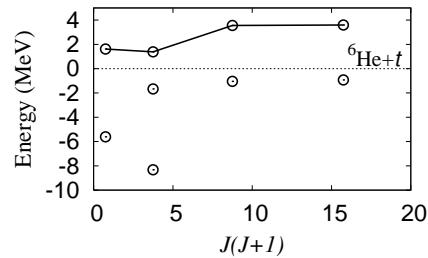}}
\caption{Energy levels of negative-parity states in $^9$Li calculated by the $^6$He+$t$-cluster GCM method 
with $d=1,2,\cdots, 8$ fm. 
Energies are measured from the $^6$He+$t$ threshold energy.
Open circles connected by the solid line are 
$K^\pi=1/2^-$ band members.}
\label{fig:li9spe}
\end{figure}

\begin{figure}[th]
\epsfxsize=5 cm
\centerline{\epsffile{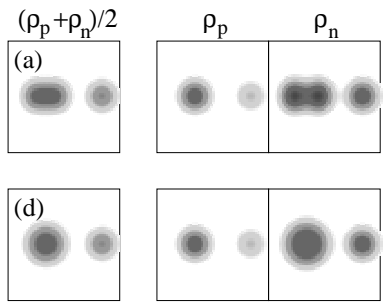}}
\caption{Density distributions of the BB $^6$He+$t$-cluster wave functions
with the configuration $(a)$ and $(d)$ at $d=5$ fm, 
$\Phi_{(a)}(d=5 \text{ fm})$ and $\Phi_{(d)}(d=5 \text{ fm})$.
Distributions of the matter, proton and neutron density are shown left, middle and right, respectively.}
\label{fig:he6-t-density}
\end{figure}

We superpose the $^6$He+$t$ cluster wave functions with $d=1,2,\cdots, 8$ fm and obtain 
energy levels of $^9$Li. 
The calculated energy and $^6$He+$t$ threshold energy are
 $-34.3$ and $-26.0$ MeV, respectively. Though the calculations overestimate
the experimental energy ($-45.3$ MeV) and the threshold energy ($-37.8$ MeV), 
they reproduce well the $^9$Li energy relative to the $^6$He+$t$ threshold.
The energy levels measured from the $^6$He+$t$ threshold energy are 
shown in Fig.~\ref{fig:li9spe}. 
The developed $^6$He+$t$ cluster states are suggested at the energy region
a few MeV higher than the $^6$He+$t$ threshold energy. 
The $1/2^-_2$, $3/2^-_3$, $5/2^-_2$, and $7/2^-_2$ states are considered to be members of 
a $K^\pi=1/2^-$ band which shows rather strong in-band $E2$ transitions 
due to the developed cluster structure (see Table \ref{tab:be2-li9}).
The intrinsic wave functions of these states contain 
dominant components of the configuration $(d)$ at $d=5$ fm, and also 
significant components of the configuration $(a)$ at $d=5$ fm. The overlap of the $3/2^-_3$ state 
with $P^{J=3/2,-}_{M,K=1/2}\Phi_{(d)}(d=5 \text{ fm})$ is 65\% and, that 
with $P^{J=3/2,-}_{M,K=1/2}\Phi_{(a)}(d=5 \text{ fm})$ is 45 \%.
As shown in Fig.~\ref{fig:he6-t-density} for the density distributions of the intrinsic wave functions, 
$\Phi_{(a)}(d=5 \text{ fm})$ and $\Phi_{(d)}(d=5 \text{ fm})$,
the configuration $(a)$ has the strong coupling feature where the $t$ cluster is sitting 
on the head of the deformed $^6$He cluster showing the elongate neutron structure,
while the configuration (d) shows a characteristic of the weak coupling feature.
Thus, the structure of the $K^\pi=1/2^-$ band is regarded as the 
$^6$He+$t$ cluster structure with the intermediate feature 
between the strong coupling and the weak coupling regimes.

As we increase the model space by adding basis wave functions with larger $d$ values,  
the energies of these states above the $^6$He+$t$ cluster threshold decrease 
because of coupling with continuum states. It means that
the GCM calculation within the $d=1,2,\cdots, 8$ fm  model space corresponds to 
a bound state approximation. 
We will discuss the stability of these resonance states later. 

\begin{figure}[th]
\epsfxsize=6 cm
\centerline{\epsffile{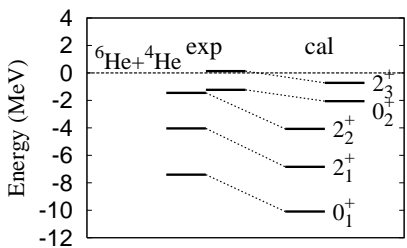}}
\caption{Energy levels of positive-parity states in $^{10}$Be obtained by the $^6$He+$^4$He-cluster GCM calculations
with $d=1,2,\cdots, 8$ fm. The experimental energy levels for the corresponding states are also shown.
Energies are measured from the $^6$He+$^4$He threshold energy.}
\label{fig:be10spe}
\end{figure}

We also calculate $^{10}$Be energy levels with the $^6$He+$^4$He-cluster GCM calculations
using the parameter $d=1,2,\cdots, 8$ fm. 
The calculated energy and the $^6$He+$\alpha$
threshold energy are $-56.9$ and $-46.8$ MeV. As well as the case of $^9$Li, 
they overestimate the experimental energy ($-65.0$ MeV) and threshold energy ($-57.6$ MeV).
As shown in Fig.~\ref{fig:be10spe}, 
the calculations reasonably reproduce the experimental energy spectra for 
the $0^+_1$, $2^+_1$, $2^+_2$, $0^+_2$, and $2^+_3$ states in $^{10}$Be.
The $K^\pi=0^+_2$ band near the $^6$He+$^4$He threshold corresponds 
to the experimental $0^+_2$, and $2^+_3$ states.
The energies measured from the $^6$He+$^4$He threshold are slightly underestimated by the 
calculations. The intrinsic structure of the $K^\pi=0^+_2$ band shows a developed 
$^6$He+$^4$He-cluster structure. The $0^+_2$ state is dominated by 
the configurations $(a)$ and $(d)$ at $d=5$ fm, each of which has about 70\% overlap with the $0^+_2$ state wave function. 
The remarkable component of the configuration $(a)$ is consistent with 
the molecular orbital structure with the $\sigma^2$ molecular orbital 
configuration 
of the  $K^\pi=0^+_2$ band suggested by 
preceding works.

\begin{table}[ht]
\caption{\label{tab:be2-li9}
$E2$ transition strengths  
in $^{9}$Li. The calculated transitions having $B(E2) \ge 1.0$ e$^2$fm$^4$ are listed. 
}
\begin{center}
\begin{tabular}{c|cc|cc}
\hline
  &   	 $B(E2)$ (e$^2$fm$^4$) \\	
 \hline
  $3/2^-_1	\rightarrow 1/2^-_1$	&4.1 \\
$3/2^-_3	\rightarrow 1/2^-_2$	&31 \\
$5/2^-_2	\rightarrow 1/2^-_2$	&15 \\
$5/2^-_1	\rightarrow 3/2^-_2$	&5.6 \\
$5/2^-_2	\rightarrow 3/2^-_3$	&3.1 \\
$7/2^-_1	\rightarrow 3/2^-_1$	&1.7 \\
$7/2^-_2	\rightarrow 3/2^-_3$	&26 \\
$7/2^-_2	\rightarrow 5/2^-_2$	&1.1 \\
\hline
\end{tabular}
\end{center}
\end{table}

\subsection{$^6$He+$t$ resonances above the threshold}
In the previous subsection, we truncated the distance parameter 
as $d\le 8$ and calculated energy levels within the bound state approximation.
However, strictly speaking, the cluster states above the $^6$He+$t$ threshold energy 
can decay into the $^6$He+$t$ channel, and these $^6$He+$t$ resonance states couple with continuum states
if we adopt an enough large model space for the inter-cluster
distance. 
In fact, in the GCM calculation, energies and wave functions for the states above the threshold 
do not converge with respect to the model space of the generator coordinate $d$. 
As we increase the model space by adding basis wave functions with large $d$ values,  
the energies of the states above the $^6$He+$t$ threshold decrease 
because of coupling with continuum states. 
We here discuss how the $^6$He+$t$ cluster states in the $K^\pi=1/2^-$ band couple with continuum states. 

To see the coupling with continuum states, we analyze 
the GCM calculation obtained with a larger model space of the distance 
$d=1,2,\cdots,15$ fm by using a pseudo potential method.
We show  in Appendix \ref{app:pseudo} applicability of an analysis with the pseudo potential method to the $2^+$ resonance 
in $\alpha$-$\alpha$ system.
Usually, a pseudo potential is used in the Analytic Continuation in the Coupling Constant(ACCC) method
\cite{kukulin77,Tanaka:1999zza,Aoyama:2003az,Funaki:2005pa}
to evaluate a complex energy pole for a resonance state. 
For the ACCC method, a high accuracy of the energy levels is required. 
However, in the present case, it is difficult to apply the ACCC method 
because the level crossing is complicated and the accuracy of the calculated energies 
is not enough for the analytic continuation. Therefore, we propose an alternative way for analysis.

We superpose $^6$He+$t$ cluster wave functions with $d=1,2,\cdots,15$ fm 
in Eq.~\ref{eq:gcm}. In the diagonalization of the Hamiltonian matrix,  
we introduce a pseudo potential and add it to the original Hamiltonian, 
\begin{eqnarray}\label{eq:pseudopotential}
\tilde{H}(\delta)&=&H+\delta \times V^{\text{pseudo}}\\
V^{\text{pseudo}}&=& \sum_{i<j} v_0 \exp[ -\frac{r_{ij}^2}{a^2_0}],
\end{eqnarray}
where $v_0=-300$ MeV and $a_0=1.0$ fm are used.
By diagonalizing the norm and Hamiltonian matrices with respect to $\tilde{H}$, we obtain 
the $k$th eigen energy $E^{J_k\pm}_{^6\text{He}+t}(\delta)$ and the eigen state, 
\begin{equation}
\ket{\tilde{\Psi}^{J_k\pm}_M(\delta)}=\sum_d \sum_{\alpha,K} \tilde{c}^{(k)}_{d,\alpha, K} P^{J\pm}_{MK} \ket{\Phi_{\alpha}(d)}, 
\end{equation}
as functions of the strength $\delta$ of the pseudo potential. When $\delta=0$,
$\tilde{H}$ equals to $H$. With increase of the strength $\delta$ of the pseudo potential, 
i.e., increase of the short-range two-body attraction added artificially, 
relative energies of resonance states to the $^6$He+$t$ threshold come down and finally 
become lower than the threshold energy. It means that, when the pseudo potential
is strong enough, 
resonance states decouple from continuum states and change to bound states, 
which we call "pseudo bound states" in this paper.

\begin{figure}[th]
\epsfxsize=6 cm
\centerline{\epsffile{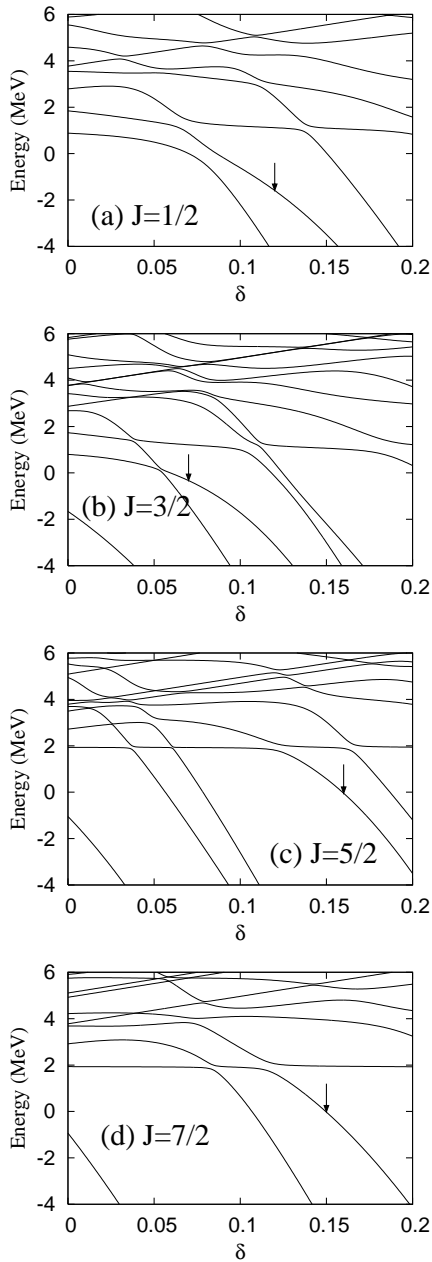}}
\caption{Energies obtained by the $^6$He+$t$-cluster GCM calculations with $d=1,2,\cdots,15$ fm 
using the pseudo potential. The energies measured from the
 $^6$He+$t$ threshold energy are plotted as a function of the strength $\delta$.}
\label{fig:he6-t.pseudo}
\end{figure}

Figure \ref{fig:he6-t.pseudo} shows energies of negative-parity states with the pseudo potential. The energies are
measured from the $^6$He+$t$ threshold energy,  
\begin{equation}
E^{J_k-}(\delta)=E^{J_k-}_{^6\text{He}+t}(\delta)-E_{^6\text{He}} (\delta) -E_{t} (\delta), 
\end{equation}
and are plotted as a function of the strength $\delta$ of the pseudo potential.
As mentioned before, the $K^\pi=1/2^-$ band is characterized by 
the significant component of the configuration $(a)$ at $d=5 \text{ fm}$ having a largely deformed neutron structure.
For an enough strength $\delta$ of the pseudo potential, 
we can easily identify the $K^\pi=1/2^-$ members 
which are specified by remarkable 
$P^{J-}_{MK}\Phi_{(a)}(d=5 \text{ fm})$ components.
The energy curves for the identified states,  
are pointed by arrows in Fig.~\ref{fig:he6-t.pseudo}.
We choose 
$\tilde{\Psi}^{1/2-}(\delta=0.12)$, $\tilde{\Psi}^{3/2-}(\delta=0.07)$, 
$\tilde{\Psi}^{5/2-}(\delta=0.16)$, and $\tilde{\Psi}^{7/2-}(\delta=0.15)$ as the pseudo bound states for the 
$K^\pi=1/2^-$ band members following the criterion that $E^{J_k-}(\delta) < 0$ and the states are decoupled from 
other states.
As $\delta$ decreases, the energies of the $^6$He+$t$ cluster states go up while 
crossing continuum states.
At $\delta=0$ for the original Hamiltonian, 
the $^6$He+$t$ resonance states couple with continuum states and they are not distinguishable
except for the $3/2^-$ state.

\begin{figure}[th]
\epsfxsize=6 cm
\centerline{\epsffile{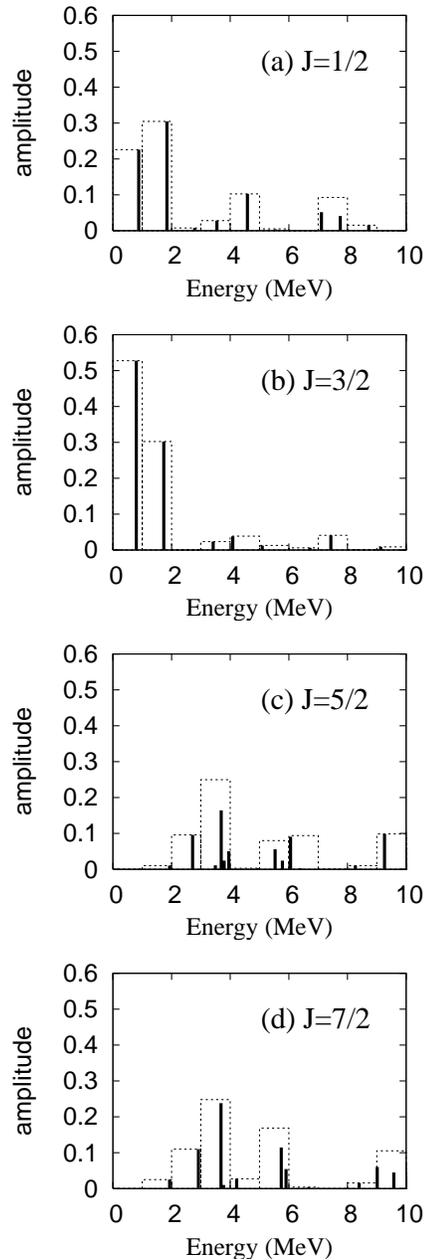}}
\caption{Solid lines: distribution of the amplitudes $|\langle \Psi^{J_k-} | \tilde{\Psi}^{J-}(\delta)\rangle |^2$ for the
$^6$He+$t$ cluster states in $^9$Li.
For the pseudo bound states $\tilde{\Psi}^{J-}(\delta)$ for the 
$^6$H+$t$ cluster states in the $K^\pi=1/2^-$ band, 
$\tilde{\Psi}^{1/2-}(\delta=0.12)$, $\tilde{\Psi}^{3/2-}(\delta=0.07)$, 
$\tilde{\Psi}^{5/2-}(\delta=0.16)$, and $\tilde{\Psi}^{7/2-}(\delta=0.15)$ are chosen.
Histograms: sum of amplitudes in each energy intervals. 
}
\label{fig:he6-t.pseudo-over}
\end{figure}

To evaluate energy spectra of the resonance states embedded in continuum states 
we analyze the overlaps of the pseudo bound states $\tilde{\Psi}^{J-}(\delta)$ 
for the $K^\pi=1/2$ band members at the finite $\delta$ values selected above 
with the wave functions $\Psi^{J_k-}$ of  energy levels at $\delta=0$.
The amplitudes $|\langle \Psi^{J_k-} | \tilde{\Psi}^{J-}(\delta)\rangle |^2$ 
indicate how the pseudo bound states $\tilde{\Psi}^{J-}(\delta)$ 
for the $^6$He+$t$ cluster states 
fragments into spectra at $\delta=0$.
The fragmentation of the amplitudes in the energy spectra may give information of resonance widths, 
because it may correspond to approximated spectra of the resonance states as shown in Appendix \ref{app:pseudo}

The distribution of the amplitudes is shown in Fig.~\ref{fig:he6-t.pseudo-over}. 
It is found that 
the amplitudes of the $1/2^-$ and $3/2^-$ states concentrate in the $0\le E\le 2$ MeV region while those of 
the $5/2^-$ and $7/2^-$ states are scattering in a wide energy region. These results suggest that 
widths of the $3/2^-$ and 
$1/2^-$ resonances may be of 1 MeV order, while  
those of the $5/2^-$ and $7/2^-$ states are expected to be larger than the $1/2^-$ and  $3/2^-$ widths.

\section{Summary}\label{sec:summary}
$^6$He+$t$ cluster states in $^9$Li were investigated by the $^6$He+$t$-cluster GCM calculation.
In the bound state approximation, the $^6$He+$t$ cluster states 
above the $^6$He+$t$ threshold energy are suggested. These states may construct a $K^\pi=1/2^-$ band. 
They have an intermediate feature between the weak coupling cluster and strong coupling cluster regimes. 
In the strong coupling regime, 
the states show a largely deformed neutron structure with the configuration similar 
to the $K^\pi=0^+_2$ band of $^{10}$Be. 

We discussed resonance features of the $^6$He+$t$ cluster states 
by analyzing the results using a pseudo potential. Amplitudes of the $1/2^-$ and $3/2^-$ states
in the energy spectra concentrate in the low-energy region, while those of the $5/2^-$ and $7/2^-$ states
fragment widely. 

\appendix
\section{Description of resonance state $^8$Be($2^+$) in an $\alpha$+$\alpha$ model}\label{app:pseudo}

\begin{figure}[tb]
\epsfxsize=6 cm
\centerline{\epsffile{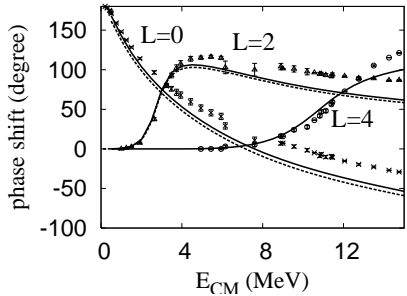}}
\caption{$\alpha$-$\alpha$ scattering phase shift calculated by using
the Volkov No.2 force with $m=0.60$. The solid and dashed lines indicate 
the phase shift obtained by $2\alpha$-cluster RGM calculations with the width parameter
$\nu=0.25$ fm$^{-2}$ and $\nu=0.235$ fm$^{-2}$, respectively. Points are experimental data.}
\label{fig:rgm}
\end{figure}

Two-body cluster states above the threshold energy are resonance states. 
When we superpose a finite number of BB wave functions without asymptotic outgoing wave boundary conditions, 
resonance states couple with continuum states
if the model space of the distance parameter $d$ is large enough. 
To see the coupling with continuum states in energy spectra, we analyze 
results of GCM calculations by using a pseudo potential method.

Let us consider here the $2^+$ resonance in $\alpha$-$\alpha$ system. 
In $^8$Be, the $2^+$  state with a width 1.51 MeV is known at 3.12 MeV from the threshold energy.
To show the applicability of the present analysis with the pseudo potential method, we apply it 
to the $\alpha$-$\alpha$ resonance.

We superpose fifteen BB $\alpha$+$\alpha$ cluster wave functions at $d=1,2,\cdots,15$ fm
in the GCM calculation.
In the diagonalization of the Hamiltonian matrix,  
we introduce the pseudo potential and add it to the original Hamiltonian $H$ as explained in 
Eq.~\ref{eq:pseudopotential}.
The effective interaction used in the present work 
is the Volkov No.2 force with $m=0.60$, which reproduces well the experimental data of   
$\alpha$-$\alpha$ scattering phase shift as shown in Fig.~\ref{fig:rgm}. 
The width parameter $\nu$ is chosen to be $\nu=0.235$ fm$^{-2}$, which is the same value as that used for the $^6$He+$t$ and 
$^6$He+$\alpha$ calculations in the present work.

\begin{figure}[th]
\epsfxsize=6 cm
\centerline{\epsffile{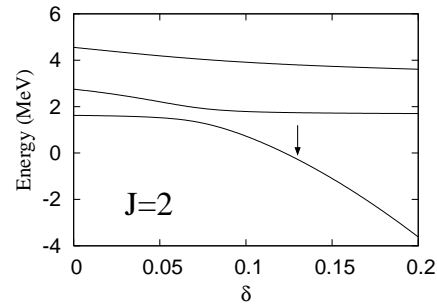}}
\caption{Energies of $2^+$ states obtained by the $2\alpha$-cluster GCM calculations with $d=1,2,\cdots,15$ fm 
using the pseudo potential. The energies measured from the
 $2\alpha$ threshold energy are plotted as a function of the strength $\delta$.}
\label{fig:he4-he4.pseudo}
\end{figure}

By diagonalizing the norm and Hamiltonian matrices with respect to $\tilde{H}$, we obtain 
the $k$th eigen energy $E^{J_k\pm}_{2\alpha}(\delta)$ and the eigen state
$\ket{\tilde{\Psi}^{J_k\pm}_M(\delta)}$ 
as functions of the strength $\delta$ of the pseudo potential. 
Figure \ref{fig:he4-he4.pseudo} shows the $2^+$ state energies $E^{2_k+}(\delta)$ 
which are measured from the threshold energy, 
\begin{equation}
E^{2_k+}(\delta)= E^{2_k+}_{2\alpha}(\delta)-2E_{\alpha}(\delta).
\end{equation} 
At $\delta=0$, the resonance $2^+$ state is obtained as the second $2^+$ state at around 3 MeV.
With increase of the strength $\delta$ of the pseudo potential, 
i.e., increase of the short-range two-body attraction added by hand, 
the energy for the resonance decreases, and it becomes lower than the threshold energy 
at $\delta=0.13$.

\begin{figure}[th]
\epsfxsize=6 cm
\centerline{\epsffile{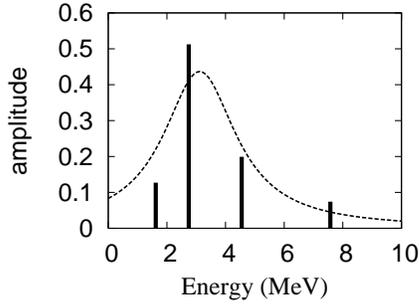}}
\caption{Solid lines: distribution of the amplitudes $|\langle \Psi^{J_k-} | \tilde{\Psi}^{J-}(\delta)\rangle |^2$
for the $2^+$ resonance in $^8$Be.
For the pseudo bound state, the wave function $\tilde{\Psi}^{J-}(\delta)$ obtained at $\delta=0.13$
is chosen. Dashed line: Breit-Wigner distribution with the experimental energy position 3.12 MeV and width 1.51 MeV.}
\label{fig:he4-he4.pseudo-over}
\end{figure}

To evaluate the energy spectra of the resonance state
we see the amplitudes of the pseudo bound state wave function for the $2^+$ state obtained at $\delta=0.13$ 
in the energy levels at $\delta=0$.
The amplitudes $|\langle \Psi^{2_k+} |\tilde{\Psi}^{2+}(\delta=0.13)\rangle |^2$ 
indicate how the $2^+$ resonance state fragments into energy spectra at $\delta=0$.
The fragmentation of the amplitudes is shown in Fig.~\ref{fig:he4-he4.pseudo-over}
compared with the Breit-Wigner distributions given by 
the experimental energy position 3.12 MeV and the width 1.51 MeV
for the $^8$Be($2^+$) state.
The calculated distribution of the amplitudes seems to correspond well to the Breit-Wigner distribution. 
This result may suggest that the amplitudes calculated by the present method using the pseudo potential
is useful to evaluate the energy and spectra of resonance states.

\section*{Acknowledgments}
The computational calculations of this work were performed by using the
supercomputers at YITP and done in Supercomputer Projects 
of High Energy Accelerator Research Organization (KEK).
This work was supported by Grant-in-Aid for Scientific Research from Japan Society for the Promotion of Science (JSPS).
It was also supported by 
the Grant-in-Aid for the Global COE Program "The Next Generation of Physics, 
Spun from Universality and Emergence" from the Ministry of Education, Culture, Sports, Science and Technology (MEXT) of Japan.

\end{document}